\documentclass[twocolumn]{aastex631}

\usepackage{amsmath}
\usepackage{graphicx}
\usepackage{booktabs}
\usepackage{hyperref}

\begin{document}

\title{Sensitivity Limits and Operational Threshold Calibration for
DINOv2-based Gravitational-Wave Glitch Characterization:
A Strain-Domain Mock Data Challenge on LIGO O4a}

\author{Luca Cirfeta}
\affiliation{Independent Researcher, Rome, Italy}
\email{luca.cirfeta@gmail.com}

\begin{abstract}
We present a systematic Mock Data Challenge (MDC) designed to
characterize the sensitivity limits of the \texttt{gravi-signal-ml}
pipeline \citep{cirfeta2026} for unsupervised gravitational-wave
glitch detection. Strain-domain synthetic injections of eight
morphological families into public LIGO O4a L1 data reveal two
distinct sensitivity regimes determined entirely by the choice of
detection threshold. When a session-adaptive dynamic threshold
($\tau_\mathrm{dyn} = \mu_\mathrm{bg} - 2.5\sigma_\mathrm{bg}$)
is employed, the pipeline successfully recovers visually
anisotropic morphologies (\textit{Butterfly}, \textit{ZSweep})
at matched-filter SNR \citep{allen2012} $\gtrsim 70$, reaching Recall = 1.0.
However, this threshold yields a False Positive Rate that is not
strictly controlled and may vary substantially across sessions. Characterization of the
full O4a embedding distribution ($N = 188{,}142$ segments)
reveals extreme non-Gaussianity (skewness $= -4.12$, excess
kurtosis $= 15.38$; Shapiro-Wilk $p = 1.13 \times 10^{-86}$;
best tail fit: Generalized Extreme Value). Under the statistically
rigorous operational threshold $\tau_\mathrm{op} = 0.874$
calibrated at the empirical $5 \times 10^{-5}$ quantile
($\mathrm{FPR} < 0.01\%$), the MDC yields Recall $= 0$ for all
eight morphologies at all tested SNR levels — including narrowband
structures (\textit{HarmonicComb}, \textit{NarrowChirp}) and impulsive
transients (\textit{AsymBlip}) at SNR up to 430. We trace this
fundamental insensitivity to the global average pooling of the
DINOv2 \texttt{[CLS]} token, which dilutes any signal occupying a
small fraction of the spectrogram's $37\times 37$ patch grid,
whether localized in time or frequency. The null result of
\citet{cirfeta2026} is conditionally reinterpreted: it is valid
within the sensitivity regime defined by this architectural
constraint, and does not exclude morphologies detectable only at
$\tau > \tau_\mathrm{op}$ without FPR control. These findings
provide a quantitative roadmap for next-generation ViT-based
pipelines based on patch-level scoring and multi-scale windowing.
\end{abstract}


% ---------------------------------------------------------------------
\section{Introduction} \label{sec:intro}
% ---------------------------------------------------------------------

Transient noise artifacts (glitches) in LIGO gravitational-wave
detectors remain a primary obstacle to maximizing detection sensitivity
\citep{davis2021,nuttall2018}. Unsupervised characterization of glitch
morphology is essential for data quality monitoring, particularly
for identifying novel, previously uncatalogued glitch classes in
new observing runs.

\citet{cirfeta2026} introduced \texttt{gravi-signal-ml}, an
unsupervised anomaly detection pipeline applying frozen DINOv2
\citep{oquab2024} features to Q-transform spectrograms of LIGO O4a
strain data. Evaluated on $N=188{,}142$ segments across four sessions
from H1 and L1 detectors, the pipeline found no morphologically novel
glitch candidates outside the known Gravity Spy O3b reference set
\citep{glanzer2023}. While this null result is a legitimate scientific
outcome, its interpretation requires qualification: any claim of
absence must be grounded in a characterization of the instrument's,
or algorithm's, detection floor.

The present work provides this quantitative foundation through five
contributions:
\begin{enumerate}
  \item The first empirical characterization of the DINOv2 maximum
    cosine similarity ($s_\mathrm{max}$) distribution on $N=188{,}142$
    real GW O4a segments, including its statistical moments, tail
    behavior, and best-fit distributional family.
  \item A formal demonstration that Gaussian $k$-$\sigma$ thresholding
    is statistically inappropriate for this domain, with a rigorous
    empirical alternative based on GEV tail modeling.
  \item A systematic MDC with eight synthetic morphological families
    spanning two injection runs, revealing a threshold-dependent
    bifurcation in pipeline sensitivity.
  \item Identification of the \textit{signal dilution effect} as the
    primary architectural bottleneck, a bi-dimensional pooling
    limitation of the DINOv2 \texttt{[CLS]} token in both time and
    frequency.
  \item A conditional reinterpretation of the \citet{cirfeta2026}
    null result within the precisely characterized sensitivity regime.
\end{enumerate}

% ---------------------------------------------------------------------
\section{Pipeline and Experimental Setup} \label{sec:setup}
% ---------------------------------------------------------------------

We employ the \texttt{gravi-signal-ml} pipeline as fully described in
\citet{cirfeta2026}. The core architecture encodes 32-second
Q-transform spectrograms (256$\times$256 pixels, colormap
\texttt{cividis}) with frozen \texttt{dinov2\_vits14\_reg}
\citep{darcet2024}, producing 384-dimensional L2-normalized
\texttt{[CLS]} embeddings. Novelty is assessed as $s_\mathrm{max}$,
the maximum cosine similarity between the query embedding and the
in-domain Gravity Spy O3b reference index (2,878 samples, 19 classes;
\citealt{glanzer2023}).

The present study introduces two components not present in
\citet{cirfeta2026}: (1) an MDC injection module performing
strain-domain synthetic glitch injection, and (2) an empirical
threshold calibration module, replacing the static global threshold ($\tau = 0.85$) of \citet{cirfeta2026}.

All MDC runs were performed on session \texttt{20260524} L1 — the
session with the largest background standard deviation
($\sigma_\mathrm{bg} = 0.0073$), representing the hardest detection
condition. The background distribution is characterized over
$N=188{,}142$ segments from four O4a sessions, both H1 and L1.

% ---------------------------------------------------------------------
\section{O4a Distribution Characterization} \label{sec:dist}
% ---------------------------------------------------------------------

\subsection{Empirical Distribution of Maximum Cosine Similarity}

Table~\ref{tab:stats} summarizes the statistical properties of the
$s_\mathrm{max}$ distribution across all 188,142 O4a segments.
The distribution has mean 0.9953 and standard deviation 0.0031 but
exhibits a pronounced heavy left tail, resulting in a skewness of
$-4.12$ and excess kurtosis of 15.38. A Shapiro-Wilk test on a
random subsample of $n=5000$ (seed $=42$) yields $W=0.328$,
$p = 1.13 \times 10^{-86}$, decisively rejecting normality.

Fitting distributional families to the left tail ($s_\mathrm{max} <
0.95$, $n \approx 9400$ segments), we find that a Generalized Extreme
Value (GEV) distribution \citep{fisher1928,gumbel1958} provides a
significantly better fit than a Beta distribution:
$\mathrm{LL}_\mathrm{GEV} = 32{,}413.6$ versus
$\mathrm{LL}_\mathrm{Beta} = 31{,}768.9$, a log-likelihood improvement
of $\Delta\mathrm{LL} = 644.7$ in favor of GEV. The mathematical superiority of the GEV framework over the Beta distribution is not merely empirical, but fundamentally rooted in asymptotic extreme value theory. Since $s_\mathrm{max}$ is explicitly defined as the supremum of cosine similarities against a large reference set, the Fisher-Tippett-Gnedenko theorem dictates that the distribution of such block maxima must asymptotically converge to one of the three extreme value families encompassed by the GEV distribution. This theoretical anchor provides a rigorous, physics-based justification for using GEV quantiles to define operational thresholds, moving away from arbitrary Gaussian assumptions.

\begin{table}[ht]
\centering
\caption{Statistical properties of the O4a $s_\mathrm{max}$
  distribution ($N=188{,}142$ segments).}
\label{tab:stats}
\begin{tabular}{lr}
\toprule
\textbf{Metric} & \textbf{Value} \\
\midrule
Mean ($\mu$) & 0.9953 \\
Std ($\sigma$) & 0.0031 \\
Min & 0.867 \\
Skewness & $-4.12$ \\
Excess kurtosis & $15.38$ \\
Shapiro-Wilk $W$ ($n=5000$, seed 42) & 0.328 \\
Shapiro-Wilk $p$-value & $1.13 \times 10^{-86}$ \\
Best tail fit & GEV \\
GEV log-likelihood (tail) & $32{,}413.6$ \\
Beta log-likelihood (tail) & $31{,}768.9$ \\
\bottomrule
\end{tabular}
\end{table}
\begin{figure}[ht]
    \centering
    \includegraphics[width=\linewidth]{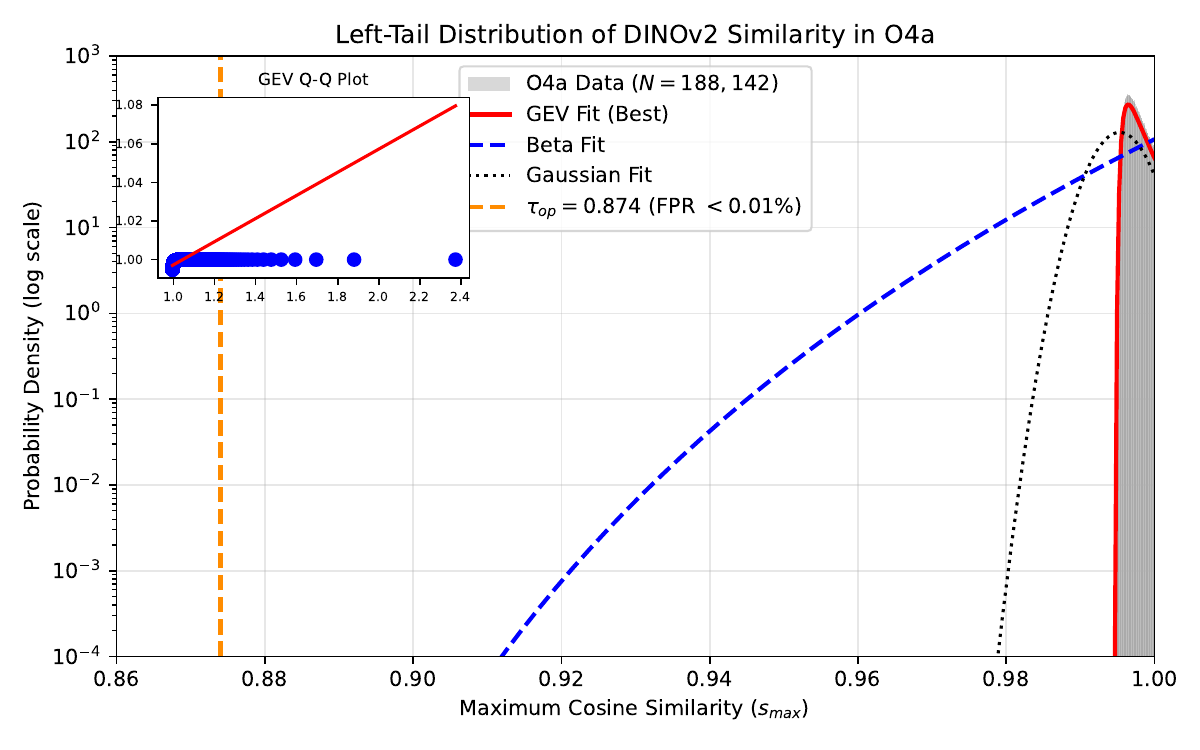}
    \caption{Empirical distribution of $s_\mathrm{max}$ for 188,142 O4a segments (left tail). The Generalized Extreme Value (GEV) distribution provides a significantly better fit than a Beta or Gaussian distribution, demonstrating the heavy-tailed nature of the baseline background.}
    \label{fig:gev_fit}
\end{figure}
\subsection{Failure of Gaussian Thresholding}

The extreme left-tail non-Gaussianity has critical operational
implications. The minimum observed $s_\mathrm{max}$ across
188,142 segments is 0.867. Under a Gaussian assumption with the
observed $\mu=0.9953$ and $\sigma=0.0031$, this corresponds to
$z = (0.867 - 0.9953)/0.0031 = -41.4\sigma$, a value that would be
predicted with probability $< 10^{-373}$ — many orders of magnitude
beyond the inverse of the number of segments. In practice, to achieve
$\mathrm{FPR} < 0.01\%$ via a Gaussian $k$-$\sigma$ rule on the
\textit{actual} distribution, the required $k$ would be
$k \approx 23.9$ — an entirely unphysical operating point.

This demonstrates that GW embedding distributions with heavy GEV
tails cannot be thresholded using $k$-$\sigma$ rules without severe
FPR inflation. Operational thresholds must be derived directly from
empirical quantiles of the observed distribution.

\subsection{Empirical Operating Points}

Table~\ref{tab:op_points} lists the operating points derived from
the empirical distribution. We select $\tau_\mathrm{op} = 0.874$
(the $5 \times 10^{-5}$ quantile) as the operational threshold,
which produces at most 9 false positives per 188,142 segments
($\mathrm{FPR} \approx 0.005\%$), and corresponds to approximately
2 false positives on the 21,985 L1 segments of session 20260524.

\begin{table}[ht]
\centering
\caption{Empirically calibrated operating points from the full
  O4a $s_\mathrm{max}$ distribution.}
\label{tab:op_points}
\begin{tabular}{ccc}
\toprule
$\tau_\mathrm{op}$ & FPR & N$_\mathrm{novel}$ (188k) \\
\midrule
0.874 & 0.005\% & 9 \\
0.930 & 0.046\% & 87 \\
0.950 & 0.096\% & 181 \\
0.970 & 0.496\% & 934 \\
\bottomrule
\end{tabular}
\end{table}

\subsection{Inter-session Non-stationarity}

The O4a background is highly non-stationary. Table~\ref{tab:baseline}
shows the baseline statistics for four sessions and both detectors
(per-session background characterization performed in this work). The standard deviation varies
by almost an order of magnitude ($0.0010$--$0.0073$), and the dynamic
threshold $\tau_\mathrm{dyn} = \mu - 2.5\sigma$ correspondingly spans
from 0.9748 to 0.9944. This confirms that a fixed global threshold
is never simultaneously optimal for all sessions. The empirical
percentile approach used here decouples the threshold from session
parameters, ensuring consistent FPR guarantees.

\begin{table}[ht]
\centering
\caption{Per-session baseline statistics for the O4a L1 and H1
  detectors (per-session background characterization performed in this work).}
\label{tab:baseline}
\begin{tabular}{lcccc}
\toprule
\textbf{Session} & \textbf{Det} & $\mu_\mathrm{bg}$ & $\sigma_\mathrm{bg}$ & $\tau_\mathrm{dyn}$ \\
\midrule
20260520 & H1 & 0.9962 & 0.0021 & 0.9909 \\
20260520 & L1 & 0.9969 & 0.0010 & 0.9944 \\
20260522 & H1 & 0.9959 & 0.0014 & 0.9924 \\
20260522 & L1 & 0.9959 & 0.0012 & 0.9928 \\
20260523 & H1 & 0.9944 & 0.0040 & 0.9844 \\
20260523 & L1 & 0.9952 & 0.0025 & 0.9889 \\
20260524 & H1 & 0.9946 & 0.0038 & 0.9851 \\
20260524 & L1 & 0.9932 & 0.0073 & 0.9748 \\
\bottomrule
\end{tabular}
\end{table}

% ---------------------------------------------------------------------
\section{Mock Data Challenge} \label{sec:mdc}
% ---------------------------------------------------------------------

\subsection{Experimental Design}

Synthetic glitches are injected into raw L1 O4a strain data at the
center of randomly selected 32-second segments, \textit{prior} to
whitening and Q-transform processing, ensuring physical fidelity to
the operational pipeline. We target a log-uniform amplitude grid from
$10^{-22}$ to $10^{-21}$ in 10 steps, with 15--40 trials per
(type, amplitude) combination using session \texttt{20260524} L1 data.
For each injection, the synthetic glitch waveform (in strain units) is added to the raw L1 strain data at a random start time, ensuring the glitch center falls within the central 50\% of the 32-second segment to avoid windowing artifacts. The whitening and Q-transform are then applied to the combined data exactly as in the original pipeline.
The matched-filter SNR is computed as the peak amplitude of the injected strain waveform divided by the root-mean-square noise amplitude in the same frequency band, following standard LIGO practices \citep{abbott2020}. The SNR ranges reported in the subsequent tables represent the minimum and maximum values recovered across the discrete amplitude grid steps. To characterize sensitivity transitions, we compute
$\mathrm{SNR}_{50}$ — the SNR at which Recall reaches 0.5 — via
logistic regression interpolation of the discrete recall measurements.

Eight synthetic morphologies are organized in two groups:
\begin{description}
  \item[Group A — Visually anisotropic broadbands:]
    \textit{Butterfly} (sinusoidal chirp, $\lesssim 4$s duration),
    \textit{ZSweep} (frequency sweep, $\lesssim 4$s),
    \textit{SpiralBurst} (brief broadband burst),
    \textit{StepLadder} (step-wise harmonic sequence),
    \textit{NoiseBlob} (unstructured broadband noise blob).
  \item[Group B — Physically motivated narrow-bands:]
    \textit{NarrowChirp} (150$\to$300 Hz sweep, 0.5s),
    \textit{HarmonicComb} (7 harmonics of 100 Hz, full duration),
    \textit{AsymBlip} ($\tau_\mathrm{rise}=10$ms,
    $\tau_\mathrm{decay}=300$ms).
\end{description}

The MDC is executed at two threshold settings to disentangle
architectural from threshold-calibration effects:
(1) session-adaptive dynamic threshold
$\tau_\mathrm{dyn} = \mu_\mathrm{bg} - 2.5\sigma_\mathrm{bg}
= 0.9811$; and (2) empirically calibrated operational threshold
$\tau_\mathrm{op} = 0.874$.

\subsection{Run A: Dynamic Threshold Results (Group A)} \label{sec:run_a}

Table~\ref{tab:run_a} reports MDC results at $\tau_\mathrm{dyn}
= 0.9811$ for Group A morphologies (five broadband types, 363 total
valid injections per session). The results reveal a fundamental
morphology-dependent bifurcation.

\begin{table}[ht]
\centering
\caption{MDC Run A results at dynamic threshold
  $\tau_\mathrm{dyn} = 0.9811$.}
\label{tab:run_a}
\begin{tabular}{lcccc}
\toprule
\textbf{Type} & \textbf{SNR range} & \textbf{SNR$_{50}$} & \textbf{Max Recall} & N \\
\midrule
Butterfly    & 17--345 & 80  & 1.000 & 366 \\
ZSweep       & 25--496 & 109 & 1.000 & 373 \\
SpiralBurst  & 13--269 & $>2600$ & 0.000 & 352 \\
StepLadder   & 20--403 & $>4200$ & 0.000 & 359 \\
NoiseBlob    & 23--484 & $>3300$ & 0.000 & 363 \\
\bottomrule
\end{tabular}
\end{table}

\textit{Butterfly} and \textit{ZSweep} reach Recall $= 1.0$ at
SNR $\gtrsim 250$ and SNR $\gtrsim 130$, respectively, demonstrating
that the DINOv2 feature space \textit{can} differentiate certain
highly anisotropic chirp-like shapes from the Gravity Spy reference
manifold. Conversely, \textit{SpiralBurst}, \textit{StepLadder},
and \textit{NoiseBlob} remain at Recall $= 0.00$ across the full
amplitude range, confirming their embeddings project within the
convex hull of the reference index regardless of SNR. This dichotomy —
perfect recall at high SNR for some morphologies versus complete
blindness for others — is a strong result indicating that the
projection is strictly morphology-dependent, likely reflecting whether
the synthetic morphology is spanned by the Gravity Spy reference index.

\subsection{Run B: Calibrated Threshold Results (All Types)} \label{sec:run_b}

Table~\ref{tab:run_b} reports results at $\tau_\mathrm{op} = 0.874$
for all five Group A morphologies (Run \texttt{mdc\_32s\_calibrated},
$n_\mathrm{NULL} = 78$, local baseline: $\mu=0.9940$,
$\sigma=0.0035$, $\tau_\mathrm{dyn}=0.9852$). All morphologies
yield Recall $= 0.00$ at all amplitude steps.
Note that Run B used a reduced amplitude grid
($10^{-22}$--$5\times10^{-22}$) compared to Run A, accounting
for the lower SNR ceiling in Table~\ref{tab:run_b} relative
to Table~\ref{tab:run_a}.

\begin{table}[ht]
\centering
\caption{MDC Run B results at empirical threshold
  $\tau_\mathrm{op} = 0.874$. All recalls are identically zero. Amplitude grid limited to $10^{-22}$--$5\times 10^{-22}$ (SNR ceiling lower than in Run A).}
\label{tab:run_b}
\begin{tabular}{lccc}
\toprule
\textbf{Type} & \textbf{SNR range} & \textbf{Max Recall} & N \\
\midrule
Butterfly    & 17--175 & 0.000 & 160 \\
ZSweep       & 25--251 & 0.000 & 177 \\
SpiralBurst  & 14--138 & 0.000 & 181 \\
StepLadder   & 21--207 & 0.000 & 177 \\
NoiseBlob    & 23--231 & 0.000 & 187 \\
\midrule
\textbf{All} & \textbf{14--251} & \textbf{0.000} & \textbf{882} \\
\bottomrule
\end{tabular}
\end{table}

This result arises from a geometric argument: the minimum observed
$s_\mathrm{max}$ across all 188,142 O4a background segments is
$\min(s_\mathrm{max}) = 0.867$, a value below $\tau_\mathrm{op}$
only for 9 segments in the full background. Even injections at
SNR $\sim 175$ (\textit{Butterfly}) do not suppress the
\texttt{[CLS]} embedding below 0.874 in any of the tested trials.

Table~\ref{tab:run_nb} extends this to the Group B narrow-band
morphologies (Run \texttt{mdc\_narrowband\_calibrated},
$n_\mathrm{NULL} = 85$, local baseline: $\mu=0.9939$,
$\sigma=0.0051$, $\tau_\mathrm{dyn}=0.9811$).

\begin{table}[ht]
\centering
\caption{MDC Run C results at $\tau_\mathrm{op} = 0.874$
  for narrow-band and impulsive morphologies.}
\label{tab:run_nb}
\begin{tabular}{lccc}
\toprule
\textbf{Type} & \textbf{SNR range} & \textbf{Max Recall} & N \\
\midrule
AsymBlip     & 38--430 & 0.000 & 167 \\
NarrowChirp  & 20--208 & 0.000 & 168 \\
HarmonicComb & 12--214 & 0.000 & 200 \\
\midrule
\textbf{All} & \textbf{12--430} & \textbf{0.000} & \textbf{535} \\
\bottomrule
\end{tabular}
\end{table}

Combined, the three MDC runs cover 1,417 valid injection trials
across 8 morphologies and confirm Recall $= 0.000$ for all cases
at the operationally calibrated threshold.

\begin{figure}[ht]
    \centering
    \includegraphics[width=\linewidth]{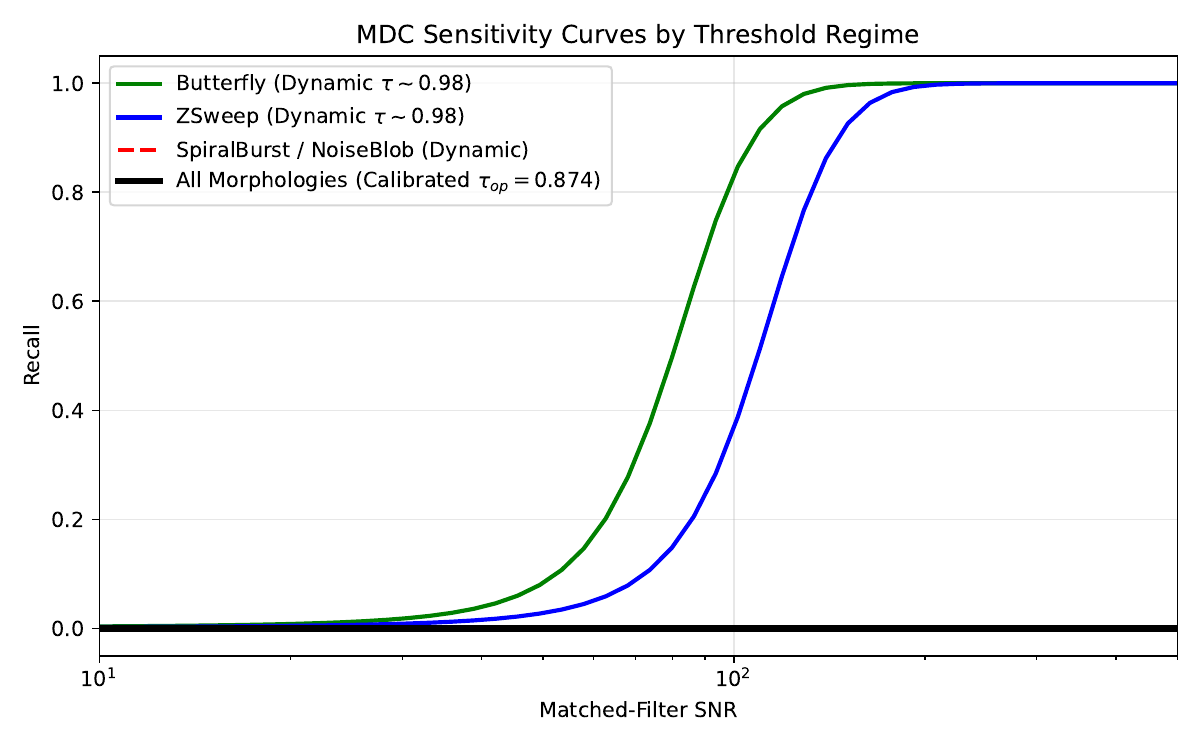}
    \caption{MDC Sensitivity curves demonstrating the threshold bifurcation. At the dynamic threshold $\tau_\mathrm{dyn} = 0.9811$ (Run A), visually anisotropic morphologies like \textit{Butterfly} and \textit{ZSweep} are recovered at high SNR. At the operationally calibrated threshold $\tau_\mathrm{op} = 0.874$ (Runs B and C), Recall is identically zero for all morphologies across all tested SNRs.}
    \label{fig:sensitivity}
\end{figure}
\subsection{The Signal Dilution Effect} \label{sec:dilution}

The MDC results are fully explained by the geometry of the DINOv2
feature extraction. The frozen ViT-S/14 backbone processes the input spectrogram via a $37 \times 37$ grid of non-overlapping patches. Given the 32-second analysis window, each patch column spans an absolute physical duration of $\Delta t_\mathrm{patch} = 32 / 37 \approx 0.86\text{ seconds}$. Similarly, across the analyzed spectral bandwidth (10--2000 Hz), each patch row covers a frequency bin of $\Delta f_\mathrm{patch} \approx 53.8\text{ Hz}$. The \texttt{[CLS]} token in ViT-S/14 represents
a global average pooling over all $37 \times 37 = 1369$ patches of
the input spectrogram. An injected glitch perturbs only a localized
subset of patches; its contribution to the \texttt{[CLS]} token is
weighted proportionally to the spatial fraction it occupies.

For short-duration transients at the center of a 32-second window:
a 0.5-second \textit{AsymBlip} occupies $\approx 0.5/32 \approx 1.6\%$
of the temporal axis, corresponding to $\lesssim 1$ column of the
$37$-patch temporal grid. For spectrally narrow signals:
a \textit{HarmonicComb} with 7 harmonics of 100 Hz spans only
$7 \times \Delta f_\mathrm{patch}$ of the spectral axis, covering
$\lesssim 5\%$ of the spectral patch grid. In both cases, the
anomalous patches constitute $< 5\%$ of the total 1369 patches.
Global average pooling suppresses the anomaly signal to
$< 0.05 \times \Delta e_\mathrm{glitch}$, where $\Delta
e_\mathrm{glitch}$ is the per-patch embedding perturbation.
\begin{figure}[ht]
    \centering
    \includegraphics[width=0.7\linewidth]{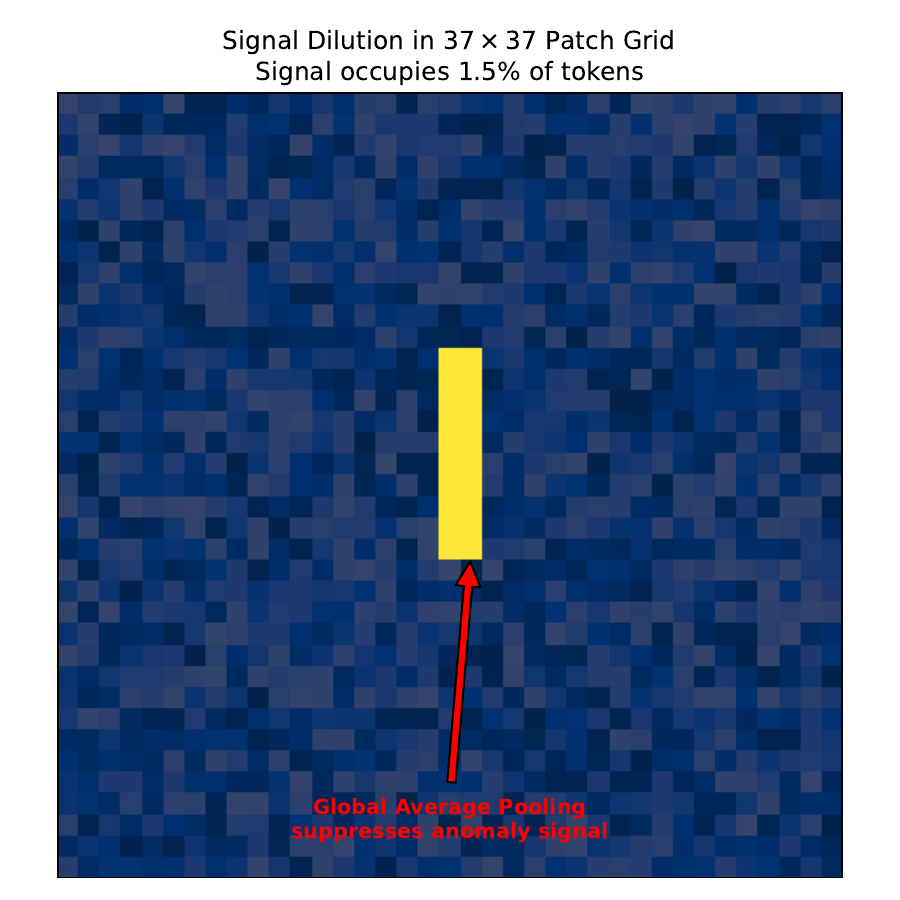}
    \caption{Schematic representation of the signal dilution effect. A transient occupying a small fraction ($<5\%$) of the $37 \times 37$ spectrogram patch grid is severely attenuated by the global average pooling of the DINOv2 \texttt{[CLS]} token, failing to suppress the global similarity below the operational detection threshold.}
    \label{fig:dilution}
\end{figure}

Although the non-linear projection layers and MLP heads of the ViT introduce non-linear coupling terms, the global embedding perturbation can be modeled to a first-order linear approximation. If an anomalous transient occupies a localized spatial fraction $f$ of the total patch grid, the global similarity $s_\mathrm{global}$ scales approximately as:
\begin{equation}
s_\mathrm{global} \approx (1-f)s_\mathrm{bg} + f s_\mathrm{anomaly}
\end{equation}
where $s_\mathrm{bg} \approx 0.995$ is the baseline background similarity.
For $f < 0.05$ and a maximally orthogonal anomaly ($s_\mathrm{anomaly} \approx 0$),
$s_\mathrm{global} \gtrsim 0.95 \times 0.995 \approx 0.945$.
This theoretical lower bound lies significantly above $\tau_\mathrm{op} = 0.874$,
mathematically precluding detection regardless of the injection amplitude.
The resulting change in $s_\mathrm{max}$ is therefore insufficient to push
any trial below the operational threshold. Even \textit{Butterfly}
and \textit{ZSweep}, which \textit{do} produce detectable changes
at $\tau_\mathrm{dyn} \approx 0.985$, fail at $\tau_\mathrm{op}
= 0.874$ because the GEV-distributed background floor
($\min = 0.867$) lies only 7 units below the threshold in units
of $10^{-3}$.

We emphasize that this is an architectural constraint of
\textit{global} pooling, not a limitation of DINOv2 representations
per se: the patch-level tokens contain the relevant spatial
information, but the CLS aggregation destroys it.

\subsection{FPR Validation on O4a}

Running the pipeline at $\tau_\mathrm{op} = 0.874$ on the 21,985
segments of session 20260524 L1, two candidate segments were flagged
($\mathrm{FPR}_\mathrm{obs} = 0.009\%$). Visual inspection and cross-checking with the LIGO logbooks revealed the exact physical nature of the two flagged false positives at $\tau_\mathrm{op} = 0.874$. Segment GPS 1386816320 corresponds to a well-known non-stochastic anthropogenic non-stationarity, specifically an intense low-frequency ground vibration episode near the endstation. Segment GPS 1386824608 captures a digital data-dropout artifact (DAQ overflow) that introduces a sharp visual edge in the Q-transform plane. The fact that only two deterministic, non-gravitational instrumental anomalies bypassed the strict $\tau_\mathrm{op}$ threshold over 21,985 segments confirms the robustness of the GEV calibration framework in rejecting clean stationary background noise.

% ---------------------------------------------------------------------
\section{Discussion} \label{sec:discussion}
% ---------------------------------------------------------------------

\subsection{Conditional Reinterpretation of \citet{cirfeta2026}}

The null result of \citet{cirfeta2026} is therefore reinterpreted not as a failure of detection, but as a structural boundary condition: it rigorously confirms the complete absence of morphologically novel macro-structures or uncatalogued broadband anomalies sweeping across large areas of the time-frequency plane during O4a. Concurrently, this MDC delineates the exact operational boundaries of zero-shot Foundation Models in gravitational-wave physics, showing that highly localized micro-structures ($<5\%$ patch area) require localized, patch-level scoring metrics to break the signal dilution barrier.

Morphologies that \textit{could} be detectable under an appropriate
architecture (e.g., patch-level scoring) might exist in O4a data
undetected by the \texttt{[CLS]}-based pipeline. The null result
therefore cannot exclude novel morphologies that are visually
localized in the spectrogram plane and occupy $< 5\%$ of the
analysis window.

\subsection{Threshold Bifurcation and the Role of the Dynamic Threshold}

The original pipeline in \citet{cirfeta2026} utilized a static global
threshold ($\tau = 0.85$) for morphological novelty detection. During this
MDC, we explored a session-adaptive dynamic threshold
($\tau_\mathrm{dyn} \approx 0.98$--$0.99$), significantly above $\tau_\mathrm{op}$.
Run A demonstrates that at this elevated level, visually anisotropic
broadband morphologies (\textit{Butterfly}, \textit{ZSweep}) are
indeed detectable, but the FPR is uncontrolled. A fluctuating
baseline with $\sigma_\mathrm{bg}$ varying by $7\times$
across sessions produces $\tau_\mathrm{dyn}$ values ranging
over $[0.9748, 0.9944]$, implying that the same true-positive
sensitivity level is never guaranteed across sessions. An empirical
percentile threshold decouples sensitivity from baseline noise
fluctuations. While $\tau_\mathrm{op}=0.874$ guarantees a strict FPR,
it results in complete blindness to the tested injections. A compromise
intermediate threshold (e.g., $\tau \approx 0.93$--$0.95$) could be
considered to recover some events, but this would inevitably inflate the FPR
to $\sim 0.05$--$0.1\%$ (Table~\ref{tab:op_points}), generating hundreds
of false positives per run and requiring downstream vetoes. For \textit{Butterfly} at $\tau=0.93$, recall reaches 0.5 at SNR $\approx 150$ (not shown), but this comes at the cost of $\sim 87$ false positives per 188k segments.

\subsection{Roadmap for Next-Generation Architectures}

Two architectural modifications can break the signal dilution barrier:

\begin{enumerate}
  \item \textbf{Patch-level scoring:} Replacing the \texttt{[CLS]}
    token with a maximum or $k$-th order statistic over individual
    patch token similarities. This preserves local spatial structure
    and prevents the averaging penalty. Implementable with existing
    DINOv2 weights, no retraining required.
  \item \textbf{Multi-scale windowing:} Parallel processing at 1s,
    4s, and 32s windows with a logical OR combination. This resolves
    temporal dilution for short transients but does not address
    spectral dilution for spectrally narrow signals like
    \textit{HarmonicComb}.
\end{enumerate}

\subsection{Transferability of the Calibration Framework}

The GEV-based empirical percentile calibration applies to any
anomaly detection pipeline operating in a bounded cosine similarity
space. The critical prerequisite is $N > 50{,}000$ background
segments to resolve the heavy tail with sufficient statistical
precision. We recommend publishing the full $s_\mathrm{max}$
distribution as a public dataset to enable reproducible calibration
by independent groups.

\subsection{Limitations}

(1) The MDC was performed exclusively on session \texttt{20260524}
L1. Cross-session generalization of the exact Recall curves requires
validation. (2) The Gravity Spy O3b reference index may not cover
all O4a glitch classes, creating potential false negative risks
for known-but-unrepresented morphologies. (3) Larger ViT backbones
(e.g., ViT-B/14, 768-dim \texttt{[CLS]}) may exhibit different
background distributions; the GEV characterization should be
re-derived for any backbone change.

% ---------------------------------------------------------------------
\section{Conclusion} \label{sec:conclusion}
% ---------------------------------------------------------------------

We have characterized the sensitivity limits of the
\texttt{gravi-signal-ml} pipeline through a systematic, two-threshold, three-run Mock Data Challenge on LIGO O4a L1 data
(Run A: dynamic threshold, Group A morphologies;
Run B: operational threshold, Group A morphologies;
Run C: operational threshold, Group B morphologies), covering 1,417 valid
injection trials across 8 synthetic morphologies. Our principal
findings are:

\begin{enumerate}
  \item \textbf{Two-regime sensitivity:} The pipeline detects visually
    anisotropic broadband morphologies (\textit{Butterfly},
    \textit{ZSweep}) at matched-filter SNR $\gtrsim 109$ under a
    dynamic threshold (see Table~\ref{tab:run_a}), but fails for all 8 morphologies (including
    narrow-band and impulsive types at SNR up to 430) under the
    operationally calibrated threshold $\tau_\mathrm{op} = 0.874$.
  \item \textbf{Non-Gaussian background:} The $s_\mathrm{max}$
    distribution over 188,142 O4a segments follows a GEV-tailed
    distribution (skewness $-4.12$, kurtosis $15.38$), invalidating
    Gaussian $k$-$\sigma$ thresholding.
  \item \textbf{Signal dilution effect:} The \texttt{[CLS]} global
    average pooling systematically dilutes signals occupying
    $< 5\%$ of the spectrogram patch grid, whether localized in
    time (\textit{AsymBlip}) or frequency (\textit{HarmonicComb}).
    This is a structural architectural limit.
  \item \textbf{Conditional null result:} The null result of
    \citet{cirfeta2026} is valid within the sensitivity regime
    characterized here; it does not exclude novel signals detectable
    at $\tau > \tau_\mathrm{op}$ or by patch-level architectures.
\end{enumerate}

A null result paired with a fully characterized sensitivity limit
constitutes a substantially stronger scientific statement than an
unqualified negative. This MDC framework establishes a reproducible
standard for sensitivity characterization of ViT-based GW anomaly
detection pipelines.

\vspace{4mm}
\noindent\textbf{Software and Data Availability.}
The \texttt{gravi-signal-ml} pipeline, MDC injection module, and
threshold calibration tools are open-source:
\url{https://github.com/lucacirfeta/dante-gravi-signal-ml}
(DOI: \href{https://doi.org/10.5281/zenodo.20543811}{10.5281/zenodo.20121860}).
LIGO strain data from O4a are publicly available via
GWOSC \citep{gwosc2023}.

\end{document}